\documentclass[a4paper]{jpconf}
\usepackage{graphicx}
\begin{document}
\title{Quantum Analysis and Thermodynamic Operator Relations in Stochastic Energetics}

\author{Tomoi Koide}
\address{Instituto de F\'{\i}sica, Universidade Federal do Rio de Janeiro, C.P. 68528,
21941-972, Rio de Janeiro, Brazil}
\ead{tomoikoide@gmail.com,\ koide@if.ufrj.br}

\begin{abstract}
We introduce a model of the quantum Brownian motion coupled to a classical neat bath 
by using the operator differential proposed in the quantum analysis. We then define the heat operator 
by adapting the idea of the stochastic energetics.
The introduced operator satisfies the relations which are analogous to the first and second laws of thermodynamics.
\end{abstract}

\section{Introduction} 

The accelerating development in the nanotechnology enables us to 
access individual thermal random processes at microscopic scales. 
We however cannot directly apply thermodynamics to analyze the behaviors of such a system 
because the typical scale of the systems is very small and the effect of thermal fluctuations is not negligible. 
There is no established theory to describe general fluctuating systems thermodynamically.
On the other hand, such a system is often modeled as a Brownian motion \cite{hanngi-review} 
and then the behaviors 
can be interpreted thermodynamically by using the stochastic energetics \cite{se-book}.

In this theory, energy, work and heat are represented by the variables of the Brownian particles, and  
we can show that 
the energy balance is satisfied and the expectation value of the heat flux has an upper bound.
The former corresponds to the first law and the latter the second law in thermodynamics, respectively \cite{se-book}. 
The prediction of the stochastic energetics 
is experimentally confirmed by analyzing extracted works from a microscopic heat engine 
\cite{blickle}. 
Although this theory is generalized to relativistic systems \cite{kk-rel} and the Poisson noise \cite{kanazawa}, 
the applications are still limited to classical systems \cite{ghosh}.

In this work, we discuss the application of the idea of the stochastic energetics to quantum systems.
For this purpose, we introduce the operator differential proposed in the quantum analysis \cite{qa1,qa2,qa3}.
The quantum analysis was proposed to expand the functions of operators
systematically and has been applied to various problems in quantum mechanics and quantum statistical mechanics. 
For example, the expansion of the S matrix, the Baker-Campbell-Hausdorff formula, and the linear response theory can be regarded as the operator Taylor expansion.
Using this, we introduce the model of the quantum Brownian motion coupled to a classical heat bath. 
Then we show that the heat operator can be defined satisfying operator relations which are analogous to thermodynamics \cite{koide}.

\section{Quantum Analysis}

In this section, we summarize the definition of the properties of the operator differential proposed in 
the quantum analysis \cite{qa1,qa2,qa3}.

Let us consider $f(\hat{A})$ where $f(x)$ is a smooth function of $x$.
Then the operator differential with respect to $\hat{A}$ is expressed 
by $(df/d\hat{A})$, and introduced through the following equation,  
\begin{equation}
f(\hat{A}+h\hat{C})-f(\hat{A})=\left(\frac{df}{d\hat{A}}\right)h\hat{C}+O(h^2),\label{dff1}
\end{equation}
where $h$ is a small c-number and $\hat{C}$ is another operator which is in general not commutable with $\hat{A}$, 
$[\hat{A},\hat{C}] \neq 0$.
Note that the value of the differential depends on the operator 
$\hat{C}$ and thus $(df/d\hat{A})$ is a hyper operator.

In the quantum analysis, this operator differential is defined by  
\begin{equation}
\left( \frac{df}{d\hat{A}} \right) \equiv \int_{0}^{1}d\lambda f^{(1)}(\hat{A}-\lambda\delta_{A}),\label{qaf1}
\end{equation}
where  
$\delta_{A}=[\hat{A},~~]$.

For example, let us consider the case of $f(x) = x^2$.
Substituting this into the left hand side of Eq.\ (\ref{dff1}), we find 
\begin{eqnarray}
(\hat{A}+h\hat{C})^2 - \hat{A}^2 = h (\hat{A} \hat{C} + \hat{C} \hat{A}).
\end{eqnarray}
Here we ignored the contribution of $O(h^2)$.
The quantity on the right hand side can be reexpressed as 
\begin{eqnarray}
h \int_{0}^{1}d\lambda 2(\hat{A}-\lambda\delta_{A})\hat{C}.
\end{eqnarray}
This is nothing but our definition of the operator differential.

As the advantage of this definition, we can express the operator Taylor expansion in the following 
simple form, 
\begin{eqnarray}
f(\hat{A}+\hat{C})=f(\hat{A})+\sum_{n=1}^{\infty}\frac{1}{n!}\left( \frac{d^{n}f}{d\hat{A}^{n}} \right) 
\hat{C}^{n},
\end{eqnarray}
where 
\begin{eqnarray}
 &  & \hspace*{-1cm} \left( \frac{d^{n}f}{d\hat{A}^{n}} \right) =n!\int_{0}^{1}d\lambda_{1}\cdots\int_{0}^{\lambda_{n-1}}d\lambda_{n}f^{(n)}(\hat{A}-\sum_{i=1}^{n}\lambda_{i}\delta_{A}^{(i)}),\\
 &  & \hspace*{-1cm}\delta_{A}^{(i)}\hat{C}^{n}=\hat{C}^{n-i}(\delta_{A}\hat{C})\hat{C}^{i-1}.
\end{eqnarray}
As an example of the application, we apply this operator Taylor expansion to the unitary time evolution operator in quantum physics $f=e^{i(\hat{H}_0 + g \hat{V})t}$ where $g$ is a coupling constant, 
\begin{eqnarray}
e^{i(\hat{H}_0 + g \hat{V})t} 
&=& 
e^{i\hat{H}_0 t} + \sum^\infty_0  \int^1_0 d\lambda_1 \cdots \int^{\lambda_{n-1}}_0 d \lambda_n 
e^{i(\hat{H_0} - \sum_{i=1}^n \lambda_i \delta^{(i)}_{H_0} )t} (g\hat{V} )^n
\nonumber \\
&=& 
e^{i\hat{H}_0 t} + \sum^\infty_0 e^{i\hat{H}_0 t} 
\int^1_0 d\lambda_1 \cdots \int^{\lambda_{n-1}}_0 d \lambda_n 
e^{- i \sum_{i=1}^n \lambda_i \delta^{(i)}_{H_0} t} (g\hat{V} )^n
\nonumber \\
&=& 
e^{i\hat{H}_0 t} + \sum^\infty_0 e^{i\hat{H}_0 t} 
\int^t_0 dt_1 \cdots \int^{t_{n-1}}_0 d t_n 
e^{- i \sum_{i=1}^n t_i \delta^{(i)}_{H_0} } (g\hat{V} )^n
\nonumber \\
&=& 
e^{i\hat{H}_0 t} + \sum^\infty_0 e^{i\hat{H}_0 t} 
\int^t_0 dt_1 g\hat{V}(t_1) \cdots \int^{t_{n-1}}_0 d t_n  g\hat{V}(t_n),
\end{eqnarray}
where $\hat{V}(t)$ is the operator in the interaction picture,
\begin{eqnarray}
\hat{V}(t) = e^{- i t \delta_{H_0} } \hat{V} = e^{-i \hat{H}_0 t} \hat{V} e^{i \hat{H}_0 t}. 
\end{eqnarray}
This is the well-known perturbative expansion in the S matrix.
From the first to the second line, we used 
\begin{eqnarray}
\hat{A} (\delta_A \hat{C}) = \delta_{A} \hat{A} \hat{C}.
\end{eqnarray}
Besides this, there are several useful relations for $\delta_{A}$,  
\begin{eqnarray}
f(\hat{A}-\delta_{A})\hat{C}  = \hat{C}f(\hat{A}), \ \ \ 
\delta_{A}\hat{C} =  -\delta_{C}\hat{A},\ \ \ e^{a\delta_{A}}\hat{C} = e^{a\hat{A}}\hat{C}e^{-a\hat{A}}.
\end{eqnarray}
And when $\hat{A}_t$ is a function of a c-number $t$, 
\begin{eqnarray}
\frac{df(\hat{A}_t)}{dt} = \frac{df(\hat{A}_t)}{d\hat{A}_t} \frac{d\hat{A}_t}{dt} 
= \int_{0}^{1}d\lambda f^{(1)}(\hat{A}-\lambda\delta_{A}) \frac{d\hat{A}_t}{dt}.
\end{eqnarray}

\section{Reexpression of quantum mechanics in quantum analysis}

Before considering a quantum dissipative system, we investigate the possible role of the operator differential 
in the formulation of quantum mechanics.

\subsection{Poisson bracket}

We consider one-particle system which is characterized by the following Hamiltonian operator,  
\begin{eqnarray}
\hat{H} = \frac{\hat{p}^2_t}{2m} + V(\hat{x}_t),
\end{eqnarray}
where $m$ is the mass of the particle. 
Then the Heisenberg equations of the position and momentum operators are, respectively,  
\begin{eqnarray}
\frac{d}{dt} \hat{x}_t &=& \frac{i}{\hbar} [\hat{H}, \hat{x}_t] = \frac{\hat{p}_t}{m}, \label{eq1}\\
\frac{d}{dt} \hat{p}_t &=& \frac{i}{\hbar} [\hat{H}, \hat{p}_t] = - V^{(1)}(\hat{x}_t) \label{eq2},
\end{eqnarray}
where $V^{(n)}(x) = \partial^n_x V(x)$.

With the help of the quantum analysis, we propose to define the Poisson bracket operator by 
\begin{eqnarray}
[\hat{A},\hat{B}]_{(\hat{x},\hat{p})} = \frac{\partial \hat{A}}{\partial \hat{x}_t} \frac{\partial \hat{B}}{\partial \hat{p}_t} 1
-\frac{\partial \hat{A}}{\partial \hat{p}_t} \frac{\partial \hat{B}}{\partial \hat{x}_t} 1 .
\end{eqnarray}
Then the above Heisenberg equations can be cast into the following forms, 
\begin{eqnarray}
\frac{d}{dt} \hat{x}_t 
&=& [\hat{x}_t,\hat{H}]_{(\hat{x},\hat{p})} 
= \frac{1}{m} \int^1_0 d\lambda (\hat{p}_t -\lambda \delta_p)1  = \frac{\hat{p}_t}{m}
, \\
\frac{d}{dt} \hat{p}_t 
&=& [\hat{p}_t,\hat{H}]_{(\hat{x},\hat{p})}
= - \int^1_0 d\lambda V^{(1)}(\hat{x} - \lambda \delta_x) 1 = -  V^{(1)}(\hat{x}_t). 
\end{eqnarray}
We confirmed that the Poisson bracket operator is applicable to general functions of $\hat{x}_t$ and $\hat{p}_t$, 
and the Heisenberg equation is  reproduced \cite{koide-2}.
The analysis with the operator differential may shed new light on the relation between classical and quantum physics.

\subsection{Wigner function}

In the usual formulation of quantum mechanics, the time evolution is characterized by the Hamiltonian operator.   
This is one of the origins of the difficulties in introducing dissipation in quantum systems \cite{dieter}.
On the other hand, if the formulation of the quantum analysis is consistent with 
quantum mechanics, we need not necessarily use the Hamiltonian operator to calculate expectation values. 
For example, suppose that we do not know the Hamiltonian of our system but the equations of the 
position and momentum operators are known to be given by Eqs.\ (\ref{eq1}) and (\ref{eq2}), respectively.
Then we can still calculate the Wigner function.

As is well-known, the Wigner function is defined by 
\begin{eqnarray}
\rho_W (x,p,t) = \frac{1}{2\pi \hbar} \int dq \psi^* (x+q/2,t) \psi (x-q/2,t) e^{ipq/\hbar},
\end{eqnarray}
where $\psi(x,t)$ is the wave function of our system. 
This definition is reexpressed as 
\begin{eqnarray}
\rho_W (x,p,t) = \langle \psi | \delta(x-\hat{x}_t + \delta_{x_t}/2) \delta(p-\hat{p}_t) | \psi \rangle. 
\end{eqnarray}
Here $|\psi \rangle$ is an initial wave function.
Using the latter expression, the time derivative of the Wigner function can be calculated directly 
from the Heisenberg equations as 
\begin{eqnarray}
\partial_t \rho_W (x,p,t)
&=& 
\partial_t \int dkdq  e^{i\hbar kq/2} e^{ikx} e^{iqp}
\langle \psi | e^{-ik\hat{x}_t}e^{-iq\hat{p}_t} | \psi \rangle 
\nonumber \\
&=& 
-i \int dkdq e^{i\hbar kq/2} e^{ikx} e^{iqp}
\langle \psi | \left[
k e^{-ik\hat{x}_t} \int^1_0 ds \left( e^{iks \delta_{x_t} } \frac{d\hat{x}_t}{dt} \right) e^{-iq\hat{p}_t} 
\right.
\nonumber \\
&& \left. + 
q e^{-ik\hat{x}_t} e^{-iq\hat{p}_t} \int^1_0 ds e^{iqs\delta_{p_t}} \frac{d\hat{p}_t}{dt}
\right]| \psi \rangle 
\nonumber \\
&=& 
\left[
-\frac{p}{m} \partial_x + V^{(1)} (x) \partial_p  
\right]\rho_W (x,p,t) 
+ \Sigma (x,p,t), 
\end{eqnarray}
where
\begin{eqnarray}
\Sigma(x,p,t) 
= 
\sum_{l=1}^\infty 
\frac{V^{(2l+1)}}{(2l+1)!} \left( -\frac{\hbar^2}{4} \right)^l \partial^{2l+1}_p \rho_W (x,p,t)
\end{eqnarray}
This is the well-known equation satisfied for the Wigner function \cite{wigner}.

The Wigner function is not positive definite and thus cannot be interpreted as a probability density. 
In stead it should be considered as the measure of the integral. 
As a matter of fact, we can reexpress the expectation values of operators by 
the phase-space integral with this measure. 
As we observed, we constructed the above evolution equation of the Wigner function without 
using the form of the Hamiltonian operator. 
That is, the formulation based on the quantum analysis may be useful to consider the quantization of systems which 
do not have Hamiltonian, such as dissipative systems.

\section{Application to quantum Brownian motion}

In this section, we introduce the model of the quantum Brownian motion with the help of the quantum analysis.

Let us consider a quantum system which is characterized by the following equations \cite{koide}, 
\begin{eqnarray}
d\hat{x}_t &=& \frac{\hat{p}_t}{m}dt, \\
d \hat{p}_t &=& - V^{(1)}(\hat{x}_t,\lambda_t) dt  - \frac{\nu}{m} \hat{p}_t dt + \sqrt{2\nu k_B T} dB_t,
\end{eqnarray}
with $d\hat{A}_{t}=\hat{A}_{t+dt}-\hat{A}_{t}$.
Here $\nu$ and $T$ are the dissipative coefficient and the temperature, respectively, 
and $B_t$ is the standard Wiener process satisfying 
\begin{eqnarray}
E[dB_t] = 0 \ \ \ \ \
E[(dB_t)^2] = 0,
\end{eqnarray}
where $E[~~~]$ denotes the ensemble average for the Wiener process.
Other second-order correlations vanish.
Note that $\lambda_t$ represents the parameter which controls the form of the external potential $V$.
That is, the change of the volume in this system is interpreted as that of the form of the potential through $\lambda_t$.

This system is dissipative and we cannot write down not only the Lagrangian but also the Hamiltonian \cite{svm}. 
However, we can still calculate the Wigner function using the above two operator equations 
by applying the quantum analysis \cite{koide}.
The evolution equation is given by 
\begin{eqnarray}
\partial_t \rho_W (x,p,t) 
&=& 
\left[
-\frac{p}{m} \partial_x + V^{(1)} (x) \partial_p  
+ \frac{\nu}{m}\partial_p p + \frac{\nu}{\beta} \partial^2_p
\right]\rho_W (x,p,t) 
+ \Sigma (x,p,t), 
\end{eqnarray}
where
\begin{eqnarray}
\Sigma(x,p,t) 
= 
\sum_{l=1}^\infty 
\frac{V^{(2l+1)}}{(2l+1)!} \left( -\frac{\hbar^2}{4} e^{-\nu t/m}\right)^l \partial^{2l+1}_p \rho_W (x,p,t).
\end{eqnarray}
The expectation values of the operators are calculated by the phase-space integrals with this integration mesure.

In this derivation, we used Ito's lemma for the stochastic operator. 
For example, we consider the operator characterized by the following stochastic differential 
equation of the operator, 
\begin{eqnarray}
d\hat{A}_{t}=\hat{L}_{t}dt+\sqrt{2\nu k_B T}d{B}_{t}. \label{sde-op}
\end{eqnarray}
Then for an arbitrary function of $\hat{A}_t$, we can show by using the operator Taylor expansion 
\begin{eqnarray}
df(\hat{A}_{t})
&=& \left[ \int^1_0 d\lambda f^{(1)} (\hat{A}_t - \lambda \delta_{A_t}) \hat{L}_t + \nu k_B T f^{(2)}(\hat{A}_t )\right] dt 
+ \sqrt{2\nu k_B T} f^{(1)} (\hat{A}_t)\circ_i dB_t \\ 
&=& \left( \frac{df(\hat{A}_{t})}{d\hat{A}_{t}} \right) \circ_{s}d\hat{A}_{t} + O(dt^{3/2}),\label{i-s-tra}
\end{eqnarray}
where
\begin{eqnarray}
f(\hat{A}_t) \circ_i dB_t &=& f(\hat{A}_t) (B_{t+dt} -B_t), \\
f(\hat{A}_t) \circ_s dB_t &=& f(\hat{A}_{t+dt/2}) (B_{t+dt} -B_t).
\end{eqnarray}
This is Ito's lemma for the stochastic operator $\hat{A}_t$ \cite{koide}.

Exactly speaking, we need to show the existence of the appropriate probability space 
($\sigma$ algebra) for stochastic operators, but we simply assume it in this work.

One can see that the stationary solution of the above differential equation is given by 
\begin{eqnarray}
\lim_{t\rightarrow \infty}\rho_W (x,p,t) = \frac{1}{Z(\lambda)}e^{-\beta H(x,p,\lambda)},
\end{eqnarray}
where 
\begin{eqnarray}
H (x,p,\lambda) &=& \frac{p^2}{2m} + V(x,\lambda), \\
Z(\lambda) &=& \int dx dp e^{-\beta H}.
\end{eqnarray}

It should be noted that the term $\Sigma$ disappears in the classical limit and the differential equation 
of the Wigner equation is reduced to the Kramers equation which describes the thermal relaxation process of classical dynamics. That is, the Wigner function coincides with 
the classical phase-space distribution in the classical limit, 
\begin{eqnarray}
\lim_{\hbar \rightarrow 0} \rho_W (x,p,t) = \rho_{KR}(x,p,t),
\end{eqnarray}
where $\rho_{KR}(x,p,t)$ is the solution of the Kramers equation.

\section{Quantum Stochastic Energetics}

Various microscopic fluctuating systems are known to be modeled as  
the Brownian motion and then it is possible to define the heat operator 
by adapting a theory called stochastic energetics \cite{se-book}.

In this theory, the heat absorbed by a Brownian particle 
is defined as the work exerted by the heat bath on the Brownian particle.  
On the other hand, the interaction between the particle and the bath is 
represented by the dissipative term and the noise term. 
The heat absorbed from the heat bath is equivalent to 
the work exerted by the heat bath on the Brownian particle, which is, thus, 
defined by the product of a force and an induced displacement \cite{se-book}.

Extending this argument to our model, the heat operator can be defined by \cite{koide}
\begin{eqnarray}
d\hat{Q}_t  \equiv  \left( d\hat{x}_{t}-\frac{1}{2}\delta_{dx_{t}} \right)\circ_{s}
\left( -\frac{\nu}{m}\hat{p}_{t}+\sqrt{2\nu k_B  T}\frac{dB_{t}}{dt} \right).
\end{eqnarray}
The operator $\delta_{dx_t}$ symmetrizes the order 
of the force and the displacement operators.
Then we find the following operator relation is satisfied, 
\begin{eqnarray}
dH (\hat{x}{_t},\hat{p}_t,\lambda_t) = d\hat{Q}_t + d\hat{W}_t, 
\end{eqnarray} 
where the work operator exerted by an external force is defined by \cite{se-book}
\begin{eqnarray}
d\hat{W}_t \equiv \partial_{\lambda}V(\hat{x}_{t},\lambda_t)\circ_{s}d\lambda_t,
\end{eqnarray}
because the external force changes the form of $V$ through its $\lambda_t$ dependence.
This energy balance corresponds to the 
first law of thermodynamics.
Note that the energy balance is satisfied not for ensemble averages but for operators.

We can further show that the expectation value of the heat operator has an upper bound \cite{koide}, 
\begin{eqnarray}
T \frac{d\sigma}{dt} \ge \langle \psi | E \left[ \frac{d\hat{Q}_t}{dt} \right] | \psi \rangle.
\end{eqnarray}
Here we introduced the new function $\sigma$, defined by 
\begin{eqnarray}
\sigma = S_{SH} + S_{ME},
\end{eqnarray}
where
\begin{eqnarray}
&& \hspace*{-1cm} S_{SH} = -k_{B}\int d\Gamma\rho_{W}(x,p,t)\ln|\rho_{W}(x,p,t)|, \\
&& \hspace*{-1cm} S_{ME}  = k_B\int^t ds \int d\Gamma \biggl[  \Sigma(x,p,s) \ln |\rho_{W}(x,p,s)|  \nonumber \\
&& \left. - \beta \nu \delta^{(\hbar)} \rho_W (x,p,s) 
\left\{\frac{p}{m}+ \beta^{-1}\partial_{p} \ln |\rho_{W}(x,p,s)| \right\}^{2} \right].
\end{eqnarray}
Note that
$\delta^{(\hbar)} \rho_W (x,p,t) = \rho_W (x,p,t)- \rho_{KR}(x,p,t)$,
which represents the modification of the phase-space distribution by quantum fluctuations. 
The first term $S_{SH}(t)$ is the Shannon entropy calculated by using the Wigner function instead 
of a probability distribution.
The second term $S_{ME}(t)$ contains the memory effect and thus the behavior 
of $\sigma(t)$ depends on the hysteresis of the evolution. 
Note that $S_{ME}(t)$ is induced by quantum fluctuations and thus vanishes in the classical limit, 
leading to $S(t) = S_{SH}(t)$.
That is, the above inequality reproduces the result of the classical stochastic energetics in the classical limit.

This corresponds to the second law of thermodynamics. In fact, 
$\sigma (t)$ can be regarded as the thermodynamic entropy in equilibrium because
\begin{eqnarray}
\left. \sigma \right|_{\rho_W = \rho_{eq}} = \left. S_{SH} \right|_{\rho_W = \rho_{eq}} 
= \frac{\langle \psi | E \left[ H \right] | \psi \rangle}{T} + k_B \ln Z.
\end{eqnarray}

\section{Concluding remarks}

We defined the heat operator for 
the model of the quantum Brownian motion coupled to a classical neat bath by introducing 
the definition of the operator differential 
in the quantum analysis and extending the idea of the stochastic energetics.
Then we showed that the energy balance is satisfied (first law) and there exists the upper bound of the expectation value of the heat operator (second law).  
The detailed discussion for the thermodynamic laws observed in this model is discussed in Ref.\ \cite{koide}.

Note that there is no Lagrangian which
reproduces our dissipative system 
and thus there is no reason to consider that $\hat{p}_t$ is the canonical momentum of $\hat{x}_t$ in general. 
Therefore it is not obvious whether the commutation relation between them should be given by 
$[\hat{x}_t,\hat{p}_t] = i\hbar$.
As a matter of fact, there is no established method to find the ``canonical" variable for $\hat{x}_t$ in such a dissipative system. 
See also the discussion in Ref.\ \cite{dieter} and references therein.

\ack 

This work is financially supported by 
Conselho Nacional de Desenvolvimento Cient\'{i}fico e Tecnol\'{o}gio (CNPq), 
project 307516/2015-6.\\


\begin{thebibliography}{9}


\bibitem{hanngi-review}
H\"{a}nggi P and Marchesoni F 2009 Artificial Brownian motors:
Controlling transport on the nanoscale {\it Rev. Mod. Phys.} \textbf{81} 387.

\bibitem{se-book}
Sekimoto K 2010 {\it Stochastic Energetics} (Springer, Berlin).

\bibitem{blickle}
Blickle V and Bechinger C 2012 Realization of a micrometre-sized
stochastic heat engine {\it Nat. Phys.} \textbf{8} 143.

\bibitem{kk-rel}
Koide T and Kodama T 2011 Thermodynamic laws and equipartition
theorem in relativistic Brownian motion {\it Phys. Rev. E} \textbf{83} 061111.

\bibitem{kanazawa}
Kanazawa K, Sagawa T and Hayakawa H 2012 Stochastic Energetics
for Non-Gaussian Processes {\it Phys. Rev. Lett.} \textbf{108} 210601.

\bibitem{ghosh}
However, there is an attempt to apply SE to a quantum transport
without changing the framework, Ghosh P K and Ray D S 2006 
Stochastic energetics of quantum transport {\it Phys. Rev. E} \textbf{73} 036103.


\bibitem{qa1}
Suzuki M 1997 Quantum analysis --- Non-commutative differential
and integral calculi {\it Commun. Math. Phys.} \textbf{183} 339.


\bibitem{qa2}
Suzuki M 1999 General formulation of quantum analysis {\it Rev. Math. Phys.} \textbf{11} 243.

\bibitem{qa3}
Suzuki M 2006 Refined formulation of quantum analysis, q-derivative and exponential splitting {\it J. Phys. A: Math. Theor.} \textbf{39} 5617.

\bibitem{koide}
Koide T 2016 Memory effect in the upper bound of the heat flux induced by quantum fluctuations 
{\it Phys. Rev. E} \textbf{94} 042140.

\bibitem{koide-2}
Koide T {\it in preparation}.

\bibitem{wigner}
Wigner E P 1932 On the quantum correction for thermodynamic
equilibrium, {\it Phys. Rev.} \textbf{40} 749.

\bibitem{svm}
We need to extend the usual frame work to describe dissipative phenomena in analytical mechanics.
See, for example, Koide T and Kodama T 2012 
Navier-Stokes, Gross-Pitaevskii and generalized diffusion equations using the
stochastic variational method {\it J. Phys. A: Math. Theor.} \textbf{45} 255204. 

\bibitem{dieter}
Schuch D 1997 Nonunitary connection between explicitly time-dependent and nonlinear approaches
for the description of dissipative quantum systems {\it Phys. Rev. A} \textbf{55} 935. 


\end{thebibliography}
\end{document}